# The First Second of a Type-II Supernova: Convection, Accretion, and Shock Propagation

H.-THOMAS JANKA[1]

Department of Astronomy and Astrophysics, University of Chicago,
5640 S. Ellis Avenue, Chicago, Illinois 60637, U.S.A.
Email: thomas@granta.uchicago.edu

AND

EWALD MÜLLER

Max-Planck-Institut für Astrophysik,
Karl-Schwarzschild-Str. 1, D-85740 Garching, Germany
Email: ewald@mpa-garching.mpg.de

ABSTRACT

One- and two-dimensional hydrodynamical simulations of the neutrino-driven supernova explosion of a $15\,M_\odot$ star are performed for the phase between the stagnation of the prompt shock and about one second after core bounce. Systematic variation of the neutrino fluxes from the neutrino sphere shows that the explosion energy, explosion time scale, initial mass of the protoneutron star, and explosive nucleosynthesis of iron-group elements depend sensitively on the strength of the neutrino heating during the first few 100 ms after shock formation. Convective overturn in the neutrino-heated region behind the shock is a crucial help for the explosion only in a narrow window of neutrino luminosities. Here powerful explosions can be obtained only in the multi-dimensional case, primarily because the overturn increases the efficiency of neutrino-energy deposition by allowing cool postshock matter to penetrate inward to the region of strongest heating, while heated gas can quickly rise outward, thus reducing its energy loss due to re-emission of neutrinos. This interpretation is supported by the different rise of the explosion energy as a function of time in 1D and 2D models. For higher core-neutrino fluxes also spherically symmetrical models yield energetic explosions, while for lower luminosities even with convection no strong explosions occur.

The cool matter in the downflows from the shock to the heating zone loses lepton number by $\nu_e$ emission, but gains energy by interactions with the core-neutrino fluxes. However, since the gas does not become very neutron-rich and its entropy increases, it is not accreted into the low-entropy, neutronized surface layer of the neutron star. Due to the absence of significant accretion while the explosions develop on a time scale of only a few 100 ms, the initial protoneutron star masses in our models are only about $1.2\,M_\odot$. Turbulent activity around the protoneutron star is a transient phenomenon; at 200–300 ms after shock formation the turbulent shell decouples from the neutrino-heated zone and moves outward behind the supernova shock. The shock shows deformation on large scales and the inhomogeneities of temperature, density, and velocity with contrasts of order unity on scales of 30° to 45° in the layer behind the shock can help to explain the anisotropies and radial mixing observed in SN 1987A. When the supernova shock reaches the entropy step of the Si-O-interface at 5700 km about 400–500 ms after bounce, the density inversion between the dense, inhomogeneous shell behind the shock and the low-density, "hot-bubble" region around the neutron star begins to steepen into a strong reverse shock. This reverse shock forms a sharp discontinuity in the neutrino-driven wind from the nascent neutron star. The deceleration of the wind expansion might trigger significant, anisotropic matter fallback to the neutron star on a time scale of several seconds.

---

[1] On leave from Max-Planck-Institut für Astrophysik, Karl-Schwarzschild-Str. 1, D-85740 Garching, Germany

*Subject headings:* supernovae: general – hydrodynamics – shock waves – convection – turbulence

## 1. INTRODUCTION

There is a variety of observational hints that large-scale mixing processes occurred and played an important role in SN 1987A even during the early phase of the explosion. X-rays (e.g., Dotani et al. 1987, Sunyaev et al. 1987) and $\gamma$-rays (e.g., Matz et al. 1988, Mahoney et al. 1988) from the decay of $^{56}$Ni and $^{56}$Co and strongly Doppler-shifted, infrared emission lines of iron (e.g., Erickson et al. 1988, Haas et al. 1990) were observed at a stage when Ni and Fe should have still been obscured by the overlying, expanding material of the progenitor star. This indicated that radioactive elements must have been mixed out with very high velocities from the place of their formation close to the center of the supernova far into the stellar mantle and envelope. The structure of the infrared emission lines was interpreted as an indication that the iron-peak elements were mixed outward macroscopically and inhomogeneously in form of about 60 to 100 identical clumps (Li et al. 1993).

The smoothness of the light curve of SN 1987A provided indirect evidence for the existence and strength of the mixing process. Mixing of hydrogen towards the center helps to explain the smooth and broad light curve maximum by the time-spread of the liberation of recombination energy (Shigeyama & Nomoto 1990). Mixing of heavy elements into the hydrogen-rich envelope homogenizes the opacity and again smooths the light curve (Arnett et al. 1989 and references therein). Moreover, the observation of a large number of fast-moving, young pulsars (Lyne & Lorimer 1994) might indicate the existence of violent, non-spherical processes in the early moments of the supernova explosion when the neutron star is formed. The large, dense explosion fragments seen on ROSAT X-ray images outside of the shock front in the Vela supernova remnant (Aschenbach et al. 1994) might possibly also originate from such early instabilities.

Numerical simulations of Rayleigh-Taylor instabilities at the composition interfaces (metal-He, He-H) in the stellar mantle and envelope after shock passage (e.g., Arnett et al. 1989, Den et al. 1990, Herant & Benz 1991) could neither explain the extent of the required mixing nor could they account for the observed high velocities and the large scales of inhomogeneities and anisotropies (Müller et al. 1991). This inspired supernova modellers to perform multi-dimensional calculations of the early phases of the supernova explosion, i.e. the core-collapse, shock-formation, and neutrino-heating phases. Spherically symmetrical models had indeed shown that convectively unstable layers are present in the collapsed stellar core after formation and propagation of the prompt shock (Burrows & Lattimer 1988). Theoretical considerations confirmed the possible importance of overturn and mixing in these layers for the evolution of the shock (Bethe 1990).

## 2. MULTI-DIMENSIONAL MODELS OF THE EXPLOSION

Herant et al. (1992) first demonstrated by a hydrodynamical simulation that strong, turbulent overturn occurs in the neutrino-heated layer outside of the protoneutron star and that this helps the stalled shock front to start re-expansion as a result of energy deposition by neutrinos. Although the existence and fast growth of these instabilities was confirmed by Janka & Müller (1993) and Müller & Janka (1994), the results of their simulations in 1D and in 2D indicated a very strong sensitivity to the conditions at the protoneutron star and to the details of the description of neutrino interactions and neutrino transport. Since the knowledge about the high-density equation of state in the nascent neutron star and about the neutrino opacities of dense matter is incomplete (Raffelt & Seckel 1994, Keil et al. 1994), the influence of a contraction of the neutron star and of the size of the neutrino fluxes on the evolution of the explosion has to be tested by systematic studies.

In the following we report the main results of a set of calculations of 1D and 2D (Newtonian) models with different core-neutrino luminosities and with varied temporal contraction of the inner boundary. The latter is placed somewhat inside the neutrino sphere and is used instead of simulating the evolution of the very dense inner core of the nascent neutron star. This gives us the freedom to set the neutrino fluxes to chosen values at the inner boundary and also enables us to follow the 2D simulations until about one second after core bounce with a reasonable number ($\mathcal{O}(10^5)$) of time steps and an acceptable computation time, i.e. several 100 h on one processor of a Cray-YMP with a grid of $400 \times 180$ zones and a highly efficient implementation of the microphysics. Doubling the angular resolution multiplies the computational load by a factor of 3–4. Our simulations are started at about 25 ms after shock formation, using an initial model that was evolved through core collapse and bounce by Bruenn (1993). The boundary luminosities of all neutrino kinds are set to values near those obtained by Bruenn (1993) and in Newtonian computations by Bruenn et al. (1995). (For details and information about the numerical treatment, see Janka & Müller 1995a).

Our approach and aims differ from those chosen by other groups who recently performed 2D simulations of supernova explosions (Herant et al. 1994, Burrows et al. 1994). These groups attempted a self-consistent treatment that includes the central, high-density part of the neutron star up to a certain stage in the evolution, typically 100–200 ms after core bounce, but they did not reveal the dependence of their results on uncertain aspects of the input physics or its numerical description.

# 3. RESULTS

## 3.1. *One-Dimensional Models*

The evolution of the stalled, prompt supernova shock in 1D models turns out to be extremely sensitive to the size of the neutrino luminosities and to the corresponding strength of neutrino heating exterior to the gain radius. Increasing the core-neutrino fluxes from low to higher values has the effect that the shock is pushed further and further out to reach a successively larger maximum radius during a phase of about 100–150 ms of slow expansion. Nevertheless, it finally recedes again to become a standing accretion shock at a much smaller radius. For a sufficiently high threshold luminosity, however, neutrino heating is strong enough to drive the shock front outwards and to cause a successful explosion. For even higher neutrino fluxes the explosion develops faster and gets more energetic. In case of our $15\,M_\odot$ star with a $1.3\,M_\odot$ iron core (Woosley et al. 1988) we find that explosions occur for electron neutrino and antineutrino luminosities in excess of $2.2 \times 10^{52}$ erg/s in case of a contracting inner boundary (to mimic the shrinking protoneutron star) but of only $1.9 \times 10^{52}$ erg/s when the radius of the inner boundary is fixed.

The transition from failure to explosion requires the neutrino luminosities to be higher than some threshold value (for a quantitative analysis, see Janka, in preparation). Yet, this is not sufficient. High neutrino-energy deposition has to be maintained for a longer period of time to ensure a successful explosion. If the decay of the neutrino fluxes is too fast, e.g., if a significant fraction of the neutrino luminosity comes from neutrino emission by spherically accreted matter, being shut off when the shock starts to expand, then the outward propagation of the shock may break down again and the model fizzles. To drive a continuous shock expansion a sufficiently high push from the neutrino-heated matter must be maintained until the material behind the shock has achieved escape velocity and does not need pressure support to make its way out (for a detailed discussion, see Janka, in preparation).

This contradicts a recent suggestion by Burrows & Goshy (1993) that the explosion can be viewed at as a global instability of the star that, once excited, inevitably leads to an explosion. The analysis by Burrows & Goshy may allow one to estimate the radius of shock stagnation. The start-up phase of the explosion, however, can hardly be described by steady-state assumptions, because the times scales of shock expansion, of neutrino cooling and heating between protoneutron star and shock, and the corresponding time scales of temperature and density changes in the postshock region are all of the same order, although they are long compared to the sound crossing time scale and may be short compared with the characteristic times of luminosity changes or variations of the mass accretion rate into the shock. In particular, due to the high sound speed and rather slow shock expansion the propagation of the shock is very sensitive to changes of the conditions in the neutrino-heated layer. A contraction of the neutron star, enhanced cooling of the gas inside the gain radius, which is a process that also accelerates the advection of matter through the gain radius and reduces the time the matter behind the shock gets heated, or even a moderate decay of the neutrino fluxes can therefore be harmful to the outward motion of the shock.

## 3.2. Two-Dimensional Models

In spherical symmetry the expansion of the neutrino-heated matter and of the shock can occur only when also the overlying material is lifted in the gravitational field of the neutron star. In the multi-dimensional case this is different. Blobs and lumps of heated matter can rise by pushing colder material aside and cold material from the region behind the shock can get closer to the zone of strongest neutrino heating to readily absorb energy. Also, when buoyancy forces drive hot matter outward, the energy loss by re-emission of neutrinos is significantly reduced. This overturn of low-entropy and high-entropy gas results in an increase of the efficiency of neutrino-energy deposition external to the gain radius and leads to explosions in 2D already for lower neutrino fluxes than in the spherically symmetrical case. Our models, however, do not show the existence of a "convective cycle" or "convective engine" (Herant et al. 1994) that transports energy from the heating region into the shock. The matter between protoneutron star and shock is subject to strong neutrino heating and cooling and our high-resolution calculations reveal a turbulent, unordered, and dynamically changing pattern of rising and sinking lumps of material with very different thermodynamical conditions and with no clear indication of inflows of cool gas and outflows of hot gas at well-defined thermodynamical states.

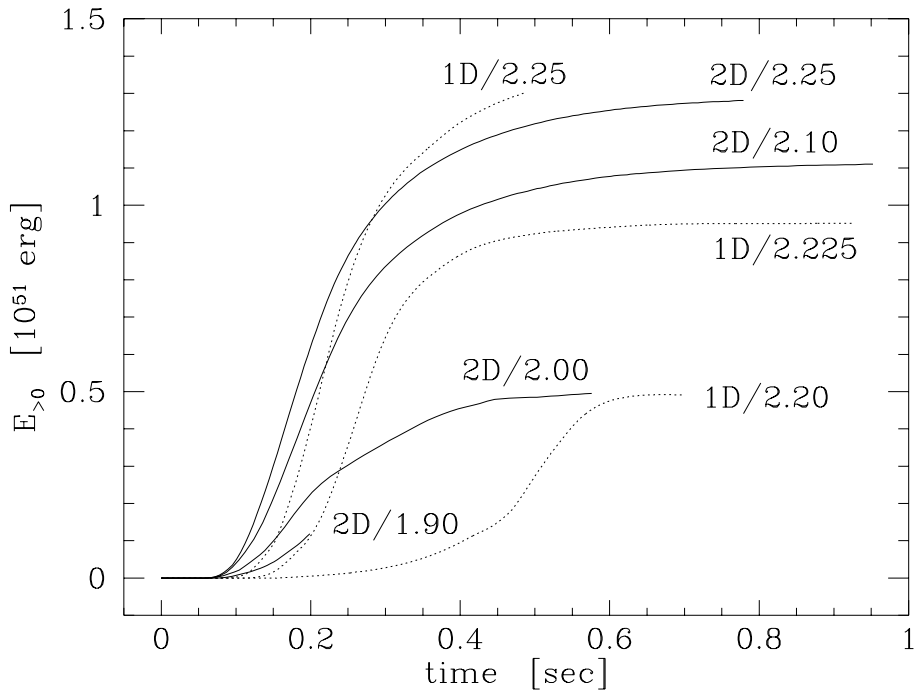

FIG. 1. Explosion energies as functions of time after the start of the simulations (about 25 ms after core bounce) for the set of exploding 1D models (dotted lines) and 2D models (solid lines). The numbers indicate the size of the initial $\nu_e$ (and $\bar{\nu}_e$) luminosities in units of $10^{52}$ erg/s.

2D models explode for core-neutrino luminosities which cannot produce explosions in 1D. There is a window of neutrino fluxes with a width of about 20% of the threshold luminosity for explosions in 1D, where convective overturn between the gain radius and the shock is a significant help for shock revival. For lower neutrino fluxes even convective overturn cannot ensure strong explosions but the explosion energy gets very low. We do not find a continuous "accumulation" (Herant et al. 1994) of energy in the convective shell until an explosion energy typical of a Type-II supernova is reached. For neutrino fluxes that cause powerful explosions already in 1D, turbulent overturn occurs but is not crucial for the explosion. In fact, in this case the effects associated with the fast rise of bubbles of heated material lead to a less vigorous start of the explosion and to the saturation of the explosion energy at a somewhat lower level (Fig. 1). The explosion energy, defined as the *net energy of the expanding matter at infinity*, does not exceed $10^{50}$ erg earlier than after about 100 ms of neutrino heating. This is the characteristic time scale of neutrinos to transfer an amount of energy to the material that is roughly equal to its gravitational binding energy and it is also the time scale that the convective overturn between gain radius and shock needs to develop to its full strength. It is not possible to determine or predict the final explosion energy of the star from a short period of only 100–200 ms after shock formation. Typically, the increase of the explosion energy with time levels off not before 400–500 ms after core bounce, followed by only a very slow increase due to the much smaller contributions of the few $10^{-3}\,M_\odot$ of matter blown away from the protoneutron star in the neutrino wind (Fig. 1). Since the wind material is heated slowly and can expand as soon as the internal energy per nucleon roughly equals its gravitational binding energy, the matter does not have a large kinetic energy at infinity.

Although the global evolution of powerful explosions in 2D, i.e., the increase of the explosion energy with time, the shock radius as a function of time, or even the amount of $^{56}$Ni produced by explosive nucleosynthesis, is not much different from energetic explosions of spherically symmetrical models, the structure of the shock and of the thick layer of expanding, dense matter behind the shock clearly show the effects of the turbulent activity. The shock is deformed on large scales and its expansion velocity into different directions varies by about 20–30%. The material behind the shock reveals large-scale inhomogeneities in density, temperature, entropy, and velocity, these quantities showing contrasts of up to a factor of 3. The typical angular scale of the largest structures is between about 30° and 45°. We do not find indications that the turbulent pattern tries to gain power on the largest possible scales and to evolve into the lowest possible mode, $l = 1$ (Herant et al. 1992, 1994). Turbulent motions are still going on in the extended, dense layer behind the shock when we stop our simulations at about 1 second after core bounce. We consider them as the origin of the anisotropies, inhomogeneities, and non-uniform distribution of radioactive elements which were observed in SN 1987A. The contrasts found behind the shock in our models are about an order of magnitude larger than the artificial perturbations that were used in the hydrodynamical simulations to trigger the growth of Rayleigh-Taylor instabilities in the stellar mantle and envelope.

### 3.3. *From Core Bounce to 1 Second*

Convective overturn outside of the protoneutron star develops within about 50–100 ms after shock formation. About 200–300 ms after bounce neutrinos have deposited a siz-able amount of energy in the material below the shock front. The turbulent layer begins to move away from the region of strongest neutrino heating and to expand outward behind the accelerating shock. This is the time when we find turbulent activity around the protoneutron star to come to an end. Our models do not give an extended phase of convection and accretion outside of the protoneutron star. The inflows of low-entropy, proton-rich gas from the postshock region towards the neutrino-heated zone are not accreted onto the protoneutron star. Although the gas loses lepton number while falling in, it does not get as neutron-rich as the material inside the gain radius. In addition, neutrino heating and mixing with the surrounding, high-entropy gas increase the entropy in the downflows. Both high electron (proton) concentration and high entropy have a stabilizing effect and prevent the penetration of the gas through the gain radius into the cooler and more neutron-rich surface layer of the protoneutron star.

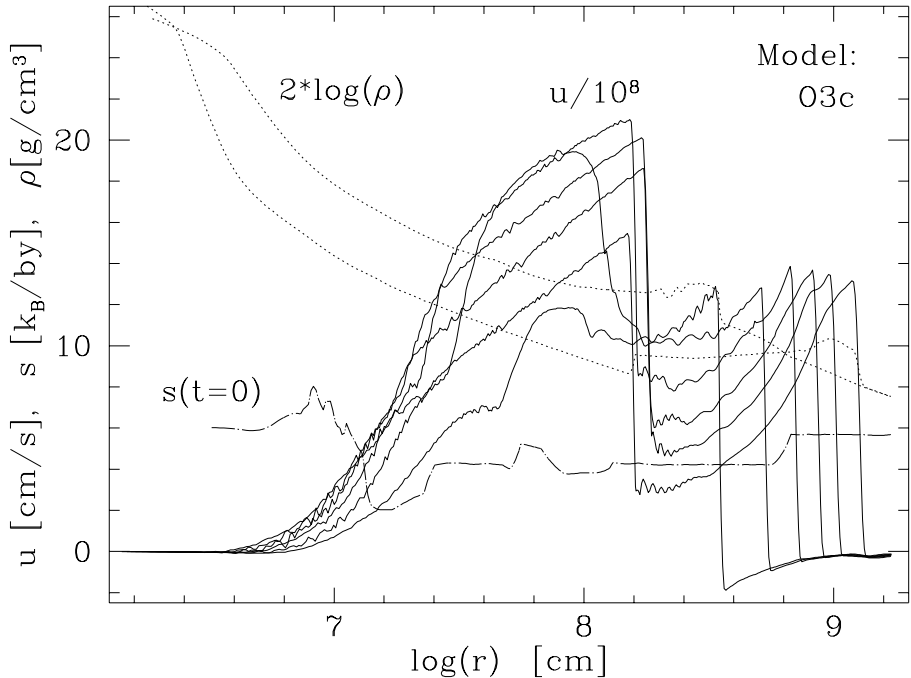

FIG. 2. Formation of the reverse shock in a 1D model with an explosion energy close to $10^{51}$ erg. The evolution of 2D models looks very similar when angle-averaged quantities are plotted, but the dense shell behind the shock is inhomogeneous and turbulent. The dash-dotted line is the entropy profile (in $k_B$/nucleon) at the start of the simulation, $t = 0$ s, about 25 ms after shock formation. The dotted lines show the density profiles at times $t = 378.4$ ms and $t = 927.2$ ms, and the solid lines display the velocity (in $10^8$ cm/s) at times $t = 378.4, 489.2, 590.7, 686.0, 777.3$, and $927.2$ ms, respectively.

At about 400–500 ms the protoneutron star has become quite compact already and the density outside of it has dropped appreciably. This indicates the formation of the high-entropy, low-density "hot-bubble" region (Bethe & Wilson 1985) and the phase of small mass loss from the nascent neutron star in the neutrino-driven wind, accompanied by slowly increasing entropies. The wind expansion accelerates as a consequence of the steepening of the density decline between the shrinking neutron star and the evacuating bubble region. The push from below creates a density inversion between the dense, slowly moving and inert shell behind the shock and the low-density hot-bubble region. At around the time the outgoing supernova shock reaches the entropy and composition discontinuity of the Si-O interface at about 5700 km, this density inversion steepens into a strong reverse shock that forms a sharp discontinuity in the neutrino wind, slowing down the wind expansion from more than $2 \times 10^9$ cm/s to a few $10^8$ cm/s (Fig. 2). We do not know yet how this reverse shock develops on a longer time scale. Since the velocities of the wind and of the layer behind the shock decrease with time it is very likely that the fallback of a significant fraction of the matter that was blown out in the neutrino wind will be triggered. Once the infall of the outer wind material is initiated and the pressure support of the gas further out vanishes, the inward acceleration might even enforce the fallback of the more slowly moving parts of the dense shell behind the supernova shock.

Fallback of a significant amount of matter, between 0.1 and $0.2\,M_\odot$, has to be postulated to solve two major problems in the current supernova models. On the one hand, due to the fast development of the explosion and the lack of an extended phase of accretion onto the collapsed stellar core, the protoneutron star formed at the center of the explosion has quite a small (initial) baryonic mass, only about $1.2\,M_\odot$ in case of our $15\,M_\odot$ star with the $1.3\,M_\odot$ iron core. On the other hand the explosive nucleosynthesis yields of iron-peak elements are incompatible with observational constraints for Type-II supernovae as deduced from terrestrial abundances and galactic evolution arguments. In case of powerful explosions with explosion energies of $1$–$1.3 \times 10^{51}$ erg about $0.2\,M_\odot$ of material are heated to temperatures above $4.5 \times 10^9$ K and are ejected behind the shock during the early phase of the explosion. Only roughly half of this matter, $0.085$–$0.1\,M_\odot$, has an electron fraction $Y_e > 0.49$ and will end up with $^{56}$Ni as the dominant nucleosynthesis product. In that respect the models seem to match the observations quite well. Yet, only some part (about $0.05\,M_\odot$) of the matter that is shock-heated to $T > 4.5 \times 10^9$ K has an electron fraction $Y_e \gtrsim 0.495$ and will end up with relative abundance yields in acceptable agreement with solar-system values. The amount of $^{56}$Ni produced in neutrino-driven explosions turns out to be correlated with the explosion energy. In case of more energetic explosions the shock is able to heat a larger mass to sufficiently high temperatures.

## 4. DISCUSSION AND CONCLUSIONS

Turbulent overturn between the zone of strongest neutrino heating and the supernova shock aids the re-expansion of the stalled shock and is able to cause powerful Type-II supernova explosions in a certain, although rather narrow, window of core neutrino fluxes where 1D models do not explode. The turbulent activity outside and close to the protoneutron star is transient and between 300 and 500 ms after core bounce the (essentially) spherically symmetrical neutrino-wind phase starts and the turbulent shell moves outward behind the expanding supernova shock. Our 2D simulations do not show a long-lasting period of convection and accretion after core bounce. Only very little of the cool, low-entropy matter that flows down from the shock front to the zone of neutrino-energy deposition is advected into the protoneutron star surface. Since the matter is proton-rich and its entropy increases quickly due to neutrino heating, it stays in the heated region to gain more energy by neutrino interactions and to start rising again. The strong, large-scale inhomogeneities and anisotropies in the expanding layer behind the outward propagating shock front will probably help to explain the effects of macroscopic mixing observed in SN 1987A and can account for moderately high recoil velocities of the neutron star (for details, see Janka & Müller 1994, 1995b).

Although the 2D models develop energetic explosions for sufficiently high neutrino luminosities and produce an amount of $^{56}$Ni that is in good agreement with observational constraints, the initial mass of the protoneutron star is clearly on the low side of the spectrum of measured neutron star masses. Moreover, the models eject about 0.1–0.2 $M_\odot$ of material with $Y_e < 0.495$, which implies an overproduction of certain elements in the iron peak by an appreciable factor compared with the nucleosynthetic composition in the solar system. The fallback of a significant fraction of this matter to the neutron star at a later stage of the evolution would ease these problems. It is possible that the reverse shock which develops in our models around the time when the supernova shock passes the entropy discontinuity of the Si-O interface will trigger this fallback on a time scale of seconds. Due to the strong inhomogeneities in the dense layer behind the shock this fallback could happen with considerable anisotropy and impart an additional kick to the neutron star (Janka & Müller 1995b).

## ACKNOWLEDGEMENTS

We are very grateful to S.W. Bruenn for providing us with the data of his core collapse calculations, which we used to construct the initial models for our simulations. This work was supported in part by the National Science Foundation under grant NSF AST 92-17969, by the National Aeronautics and Space Administration under grant NASA NAG 5-2081, and by an Otto Hahn Postdoctoral Scholarship of the Max-Planck-Society (H.-Th. J.). The computations were performed on the CRAY-YMP 4/64 of the Rechenzentrum Garching.